\pdfoutput=1

\documentclass[12pt,a4paper]{article}

\usepackage{ifthen} 
\usepackage{upgreek}
\newboolean{pdflatex}
\setboolean{pdflatex}{true} 

\newboolean{articletitles}
\setboolean{articletitles}{true} 

\newboolean{uprightparticles}
\setboolean{uprightparticles}{false} 

\def\Kbar  {\kern 0.2em\overline{\kern -0.2em \PK}{}\xspace}

\newcommand{\BF}{\ensuremath{\mathcal{B}}}

\newcommand{\Kpi}{\ensuremath{K^-\pi^+}\xspace}
\newcommand{\Kpii}{\ensuremath{K^+\pi^-}\xspace}
\newcommand{\KK}{\ensuremath{K^+K^-}\xspace}

\newcommand{\BsKstKst}{\ensuremath{B^0_s\to \Kstarz \Kstarzb}\xspace}
\newcommand{\BdKstKst}{\ensuremath{B^0\to \Kstarz \Kstarzb}\xspace}

\newcommand{\BsPhiKst}{\ensuremath{\Bs \to \phi \Kstarzb}\xspace}

\newcommand{\BdPhiKst}{\ensuremath{\Bd \to \phi \Kstarz}\xspace}
\newcommand{\BsPhiPhi}{\ensuremath{\Bs \to \phi \phi}\xspace}
\newcommand{\Bzmeson}{\ensuremath{B^{0}_{(s)}}\xspace}

\newcommand{\IP}{\ensuremath{{\rm IP}}\xspace}
\newcommand{\chisqIP}{\ensuremath{\chisq_{\IP}}\xspace}

\newcommand{\swave}{{\rm S-wave}\xspace}

\newcommand{\gaussian}{Gaussian\xspace}

\usepackage[normalem]{ulem} 
\usepackage{soul} 

\usepackage{microtype}
\usepackage{lineno}  
\usepackage{xspace} 

\usepackage{graphicx}  
\usepackage{color}
\usepackage{colortbl}
\graphicspath{{./figs/}} 

\usepackage{amsmath} 
\usepackage{amssymb}
\usepackage{amsfonts}
\usepackage{upgreek} 

\newcommand*\patchAmsMathEnvironmentForLineno[1]{%
\expandafter\let\csname old#1\expandafter\endcsname\csname #1\endcsname
\expandafter\let\csname oldend#1\expandafter\endcsname\csname
end#1\endcsname
 \renewenvironment{#1}%
   {\linenomath\csname old#1\endcsname}%
   {\csname oldend#1\endcsname\endlinenomath}%
}
\newcommand*\patchBothAmsMathEnvironmentsForLineno[1]{%
  \patchAmsMathEnvironmentForLineno{#1}%
  \patchAmsMathEnvironmentForLineno{#1*}%
}
\AtBeginDocument{%
\patchBothAmsMathEnvironmentsForLineno{equation}%
\patchBothAmsMathEnvironmentsForLineno{align}%
\patchBothAmsMathEnvironmentsForLineno{flalign}%
\patchBothAmsMathEnvironmentsForLineno{alignat}%
\patchBothAmsMathEnvironmentsForLineno{gather}%
\patchBothAmsMathEnvironmentsForLineno{multline}%
}

\usepackage{hyperref}    
\usepackage[all]{hypcap} 

\def\lhcb {\mbox{LHCb}\xspace}
\def\ux85 {\mbox{UX85}\xspace}

\def\lhc    {\mbox{LHC}\xspace}

\def\babar  {\mbox{BaBar}\xspace}
\def\belle  {\mbox{Belle}\xspace}

\ifthenelse{\boolean{uprightparticles}}%
{

 \def\Ppi         {\ensuremath{\uppi}\xspace}

 \def\Ppsi        {\ensuremath{\uppsi}\xspace}

 \def\PDelta      {\ensuremath{\Delta}\xspace}                 
 \def\PXi      {\ensuremath{\Xi}\xspace}                 
 \def\PLambda      {\ensuremath{\Lambda}\xspace}                 
 \def\PSigma      {\ensuremath{\Sigma}\xspace}                 
 \def\POmega      {\ensuremath{\Omega}\xspace}                 
 \def\PUpsilon      {\ensuremath{\Upsilon}\xspace}                 
 
 \def\PB      {\ensuremath{\mathrm{B}}\xspace}                 
                  
 \def\PD      {\ensuremath{\mathrm{D}}\xspace}

 \def\PJ      {\ensuremath{\mathrm{J}}\xspace}                 
 \def\PK      {\ensuremath{\mathrm{K}}\xspace}

 \def\Pb      {\ensuremath{\mathrm{b}}\xspace}                 
 \def\Pc      {\ensuremath{\mathrm{c}}\xspace}                 
 \def\Pd      {\ensuremath{\mathrm{d}}\xspace}

 \def\Pi      {\ensuremath{\mathrm{i}}\xspace}

 \def\Pp      {\ensuremath{\mathrm{p}}\xspace}

 \def\Ps      {\ensuremath{\mathrm{s}}\xspace}

}
{

 \def\Ppi         {\ensuremath{\pi}\xspace}

 \def\Ppsi        {\ensuremath{\psi}\xspace}                 
                  
 \mathchardef\PDelta="7101
 \mathchardef\PXi="7104
 \mathchardef\PLambda="7103
 \mathchardef\PSigma="7106
 \mathchardef\POmega="710A
 \mathchardef\PUpsilon="7107
                  
 \def\PB      {\ensuremath{B}\xspace}                 
                  
 \def\PD      {\ensuremath{D}\xspace}

 \def\PJ      {\ensuremath{J}\xspace}                 
 \def\PK      {\ensuremath{K}\xspace}

 \def\Pb      {\ensuremath{b}\xspace}                 
 \def\Pc      {\ensuremath{c}\xspace}                 
 \def\Pd      {\ensuremath{d}\xspace}

 \def\Pi      {\ensuremath{i}\xspace}

 \def\Pp      {\ensuremath{p}\xspace}

 \def\Ps      {\ensuremath{s}\xspace}

}

\def\dquark    {\ensuremath{\Pd}\xspace}

\def\squark    {\ensuremath{\Ps}\xspace}

\def\cquark    {\ensuremath{\Pc}\xspace}

\def\bquark    {\ensuremath{\Pb}\xspace}

\def\pion  {\ensuremath{\Ppi}\xspace}

\def\kaon  {\ensuremath{\PK}\xspace}
  \def\Kbar  {\kern 0.2em\overline{\kern -0.2em \PK}{}\xspace}

\def\Kz    {\ensuremath{\kaon^0}\xspace}
\def\Kzb   {\ensuremath{\Kbar^0}\xspace}
\def\KzKzb {\ensuremath{\Kz \kern -0.16em \Kzb}\xspace}
\def\Kp    {\ensuremath{\kaon^+}\xspace}
\def\Km    {\ensuremath{\kaon^-}\xspace}

\def\KpKm  {\ensuremath{\Kp \kern -0.16em \Km}\xspace}
\def\KS    {\ensuremath{\kaon^0_{\rm\scriptscriptstyle S}}\xspace} 
 
\def\Kstarz  {\ensuremath{\kaon^{*0}}\xspace}
\def\Kstarzb {\ensuremath{\Kbar^{*0}}\xspace}

\def\Kstarp  {\ensuremath{\kaon^{*+}}\xspace}

  \def\Dbar    {\kern 0.2em\overline{\kern -0.2em \PD}{}\xspace}
\def\D       {\ensuremath{\PD}\xspace}

\def\Dz      {\ensuremath{\D^0}\xspace}
\def\Dzb     {\ensuremath{\Dbar^0}\xspace}
\def\DzDzb   {\ensuremath{\Dz {\kern -0.16em \Dzb}}\xspace}
\def\Dp      {\ensuremath{\D^+}\xspace}
\def\Dm      {\ensuremath{\D^-}\xspace}

\def\DpDm    {\ensuremath{\Dp {\kern -0.16em \Dm}}\xspace}

\def\Ds      {\ensuremath{\D^+_\squark}\xspace}

\def\B       {\ensuremath{\PB}\xspace}
\def\Bbar    {\ensuremath{\kern 0.18em\overline{\kern -0.18em \PB}{}}\xspace}

\def\Bu      {\ensuremath{\B^+}\xspace}

\def\Bp      {\ensuremath{\Bu}\xspace}

\def\Bd      {\ensuremath{\B^0}\xspace}
\def\Bs      {\ensuremath{\B^0_\squark}\xspace}
\def\Bsb     {\ensuremath{\Bbar^0_\squark}\xspace}

\def\jpsi     {\ensuremath{{\PJ\mskip -3mu/\mskip -2mu\Ppsi\mskip 2mu}}\xspace}

  \def\Y#1S{\ensuremath{\PUpsilon{(#1S)}}\xspace}

\def\proton      {\ensuremath{\Pp}\xspace}
\def\antiproton  {\ensuremath{\overline \proton}\xspace}

\def\L {\ensuremath{\PLambda}\xspace}
\def\Lbar {\ensuremath{\kern 0.1em\overline{\kern -0.1em\PLambda}}\xspace}

\def\Lb      {\ensuremath{\L^0_\bquark}\xspace}
\def\Lbbar   {\ensuremath{\Lbar^0_\bquark}\xspace}
\def\Lc      {\ensuremath{\L^+_\cquark}\xspace}

\def\BF         {{\ensuremath{\cal B}\xspace}}

\def\BR         {\BF}
\newcommand{\decay}[2]{\ensuremath{#1\!\to #2}\xspace}         

\def\to                 {\ensuremath{\rightarrow}\xspace}

\def\CP                {\ensuremath{C\!P}\xspace}

\def\BdToJPsiKst  {\decay{\Bd}{\jpsi\Kstarz}}

\def\AT#1     {\ensuremath{A_{\mathrm{T}}^{#1}}\xspace}           

\def\C#1      {\ensuremath{\mathcal{C}_{#1}}\xspace}                       
\def\Cp#1     {\ensuremath{\mathcal{C}_{#1}^{'}}\xspace}                    
\def\Ceff#1   {\ensuremath{\mathcal{C}_{#1}^{\mathrm{(eff)}}}\xspace}        
\def\Cpeff#1  {\ensuremath{\mathcal{C}_{#1}^{'\mathrm{(eff)}}}\xspace}       
\def\Ope#1    {\ensuremath{\mathcal{O}_{#1}}\xspace}                       
\def\Opep#1   {\ensuremath{\mathcal{O}_{#1}^{'}}\xspace}                    

\newcommand{\tev}{\ensuremath{\mathrm{\,Te\kern -0.1em V}}\xspace}
\newcommand{\gev}{\ensuremath{\mathrm{\,Ge\kern -0.1em V}}\xspace}
\newcommand{\mev}{\ensuremath{\mathrm{\,Me\kern -0.1em V}}\xspace}
\newcommand{\kev}{\ensuremath{\mathrm{\,ke\kern -0.1em V}}\xspace}
\newcommand{\ev}{\ensuremath{\mathrm{\,e\kern -0.1em V}}\xspace}
\newcommand{\gevc}{\ensuremath{{\mathrm{\,Ge\kern -0.1em V\!/}c}}\xspace}
\newcommand{\mevc}{\ensuremath{{\mathrm{\,Me\kern -0.1em V\!/}c}}\xspace}
\newcommand{\gevcc}{\ensuremath{{\mathrm{\,Ge\kern -0.1em V\!/}c^2}}\xspace}
\newcommand{\gevgevcccc}{\ensuremath{{\mathrm{\,Ge\kern -0.1em V^2\!/}c^4}}\xspace}
\newcommand{\mevcc}{\ensuremath{{\mathrm{\,Me\kern -0.1em V\!/}c^2}}\xspace}

\def\mm   {\ensuremath{\rm \,mm}\xspace}

\def\mum  {\ensuremath{\,\upmu\rm m}\xspace}

\newcommand{\stat}{\ensuremath{\mathrm{(stat)}}\xspace}
\newcommand{\syst}{\ensuremath{\mathrm{(syst)}}\xspace}

\newcommand{\chisq}{\ensuremath{\chi^2}\xspace}

\def\gsim{{~\raise.15em\hbox{$>$}\kern-.85em
          \lower.35em\hbox{$\sim$}~}\xspace}
\def\lsim{{~\raise.15em\hbox{$<$}\kern-.85em
          \lower.35em\hbox{$\sim$}~}\xspace}

\def\pt         {\mbox{$p_{\rm T}$}\xspace}

\def\dllkpi     {\ensuremath{\mathrm{DLL}_{\kaon\pion}}\xspace}

\def\dllpk      {\ensuremath{\mathrm{DLL}_{\proton\kaon}}\xspace}

\def\evtgen     {\mbox{\textsc{EvtGen}}\xspace}
\def\pythia     {\mbox{\textsc{Pythia}}\xspace}

\def\geant      {\mbox{\textsc{Geant4}}\xspace}

\def\photos     {\mbox{\textsc{Photos}}\xspace}

\def\tell1  {TELL1\xspace}
\def\ukl1   {UKL1\xspace}

\usepackage{cite} 
\usepackage{mciteplus}

\usepackage{longtable} 
\usepackage{accents}
\newcommand*{\fancybar}{\scalebox{.4}{(}\raisebox{-1.7pt}{\bf{--}}\scalebox{.4}{)}}
\newcommand*{\brabar}[1]{\accentset{\fancybar}{#1}}
\begin{document}

\renewcommand{\thefootnote}{\fnsymbol{footnote}}
\setcounter{footnote}{1}


\begin{titlepage}
\pagenumbering{roman}

\vspace*{-1.5cm}
\centerline{\large EUROPEAN ORGANIZATION FOR NUCLEAR RESEARCH (CERN)}
\vspace*{1.5cm}
\hspace*{-0.5cm}
\begin{tabular*}{\linewidth}{lc@{\extracolsep{\fill}}r}
\ifthenelse{\boolean{pdflatex}}
{\vspace*{-2.7cm}\mbox{\!\!\!\includegraphics[width=.14\textwidth]{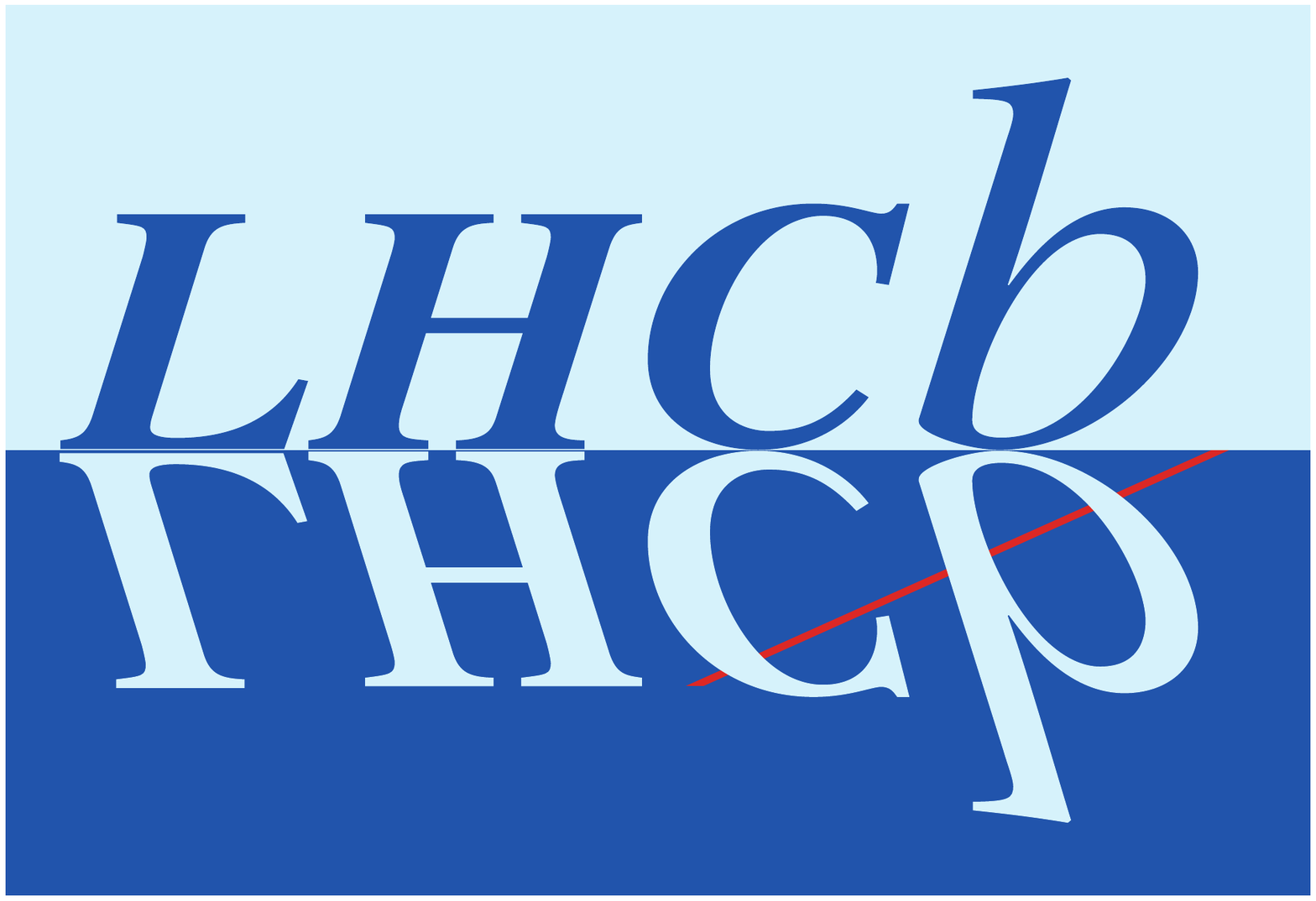}} & &}%
{\vspace*{-1.2cm}\mbox{\!\!\!\includegraphics[width=.12\textwidth]{figs/lhcb-logo.eps}} & &}%
\\
 & & CERN-PH-EP-2013-087 \\  
 & & LHCb-PAPER-2013-012 \\  
 & & 10 June 2013 \\ 
 & & \\
\end{tabular*}

\vspace*{4.0cm}

{\bf\boldmath\huge
\begin{center}
First observation of the decay \BsPhiKst
\end{center}
}

\vspace*{1.5cm}

\begin{center}
The LHCb collaboration\footnote{Authors are listed on the following pages.}
\end{center}

\vspace{\fill}

\begin{abstract}
  \noindent
The first observation of the decay \BsPhiKst is reported. The
analysis is based on a data sample corresponding to an integrated luminosity of 1.0\,fb$^{-1}$
of $pp$ collisions at $\sqrt{s} = 7\,$\tev, collected with the \lhcb detector.
A yield of ${30 \pm 6}$  ${\Bs \to (\KK)(\Kpi)}$ decays is found in the mass
windows ${1012.5 < M(\KK) < 1026.5\mevcc}$ and ${746 < M(\Kpi)< 1046\mevcc}$.
The signal yield is found to be dominated by \BsPhiKst
decays, and the corresponding branching fraction is measured to be
${\BF(\BsPhiKst) = \left(1.10 \pm 0.24\,\stat \pm 0.14\,\syst \pm 0.08\left(f_d/f_s\right)\right) \times 10^{-6}}$,
where the uncertainties are statistical, systematic and from the ratio of fragmentation fractions $f_d/f_s$
which accounts for the different production rate of \Bd and \Bs mesons. The significance of \BsPhiKst
signal is 6.1 standard deviations.
The fraction of longitudinal polarization in \BsPhiKst decays is found to be ${f_0 = 0.51 \pm 0.15\,\stat \pm 0.07\,\syst}$.
\end{abstract}

\vspace*{0.5cm}
\begin{center}
Submitted to JHEP
\end{center}

\vspace{\fill}

{\footnotesize 
\centerline{\copyright~CERN on behalf of the \lhcb collaboration, license \href{http://creativecommons.org/licenses/by/3.0/}{CC-BY-3.0}.}}
\vspace*{2mm}

\end{titlepage}


\newpage
\setcounter{page}{2}
\mbox{~}
\newpage

\centerline{\large\bf LHCb collaboration}
\begin{flushleft}
\small
R.~Aaij$^{40}$, 
C.~Abellan~Beteta$^{35,n}$, 
B.~Adeva$^{36}$, 
M.~Adinolfi$^{45}$, 
C.~Adrover$^{6}$, 
A.~Affolder$^{51}$, 
Z.~Ajaltouni$^{5}$, 
J.~Albrecht$^{9}$, 
F.~Alessio$^{37}$, 
M.~Alexander$^{50}$
% ...

{\footnotesize \it
$ ^{1}$Centro Brasileiro de Pesquisas F\'{i}sicas (CBPF), Rio de Janeiro, Brazil\\
% ...
$ ^{5}$Clermont Universit\'{e}, Universit\'{e} Blaise Pascal, CNRS/IN2P3, LPC, Clermont-Ferrand, France\\
$ ^{6}$CPPM, Aix-Marseille Universit\'{e}, CNRS/IN2P3, Marseille, France\\
% ...
$ ^{9}$Fakult\"{a}t Physik, Technische Universit\"{a}t Dortmund, Dortmund, Germany\\
% ...
$ ^{35}$Universitat de Barcelona, Barcelona, Spain\\
$ ^{36}$Universidad de Santiago de Compostela, Santiago de Compostela, Spain\\
$ ^{37}$European Organization for Nuclear Research (CERN), Geneva, Switzerland\\
% ...
$ ^{40}$Nikhef National Institute for Subatomic Physics, Amsterdam, The Netherlands\\
% ...
$ ^{45}$H.H. Wills Physics Laboratory, University of Bristol, Bristol, United Kingdom\\
% ...
$ ^{50}$School of Physics and Astronomy, University of Glasgow, Glasgow, United Kingdom\\
$ ^{51}$Oliver Lodge Laboratory, University of Liverpool, Liverpool, United Kingdom\\
% ...
$ ^{n}$LIFAELS, La Salle, Universitat Ramon Llull, Barcelona, Spain\\
% ...
}
\end{flushleft}

\cleardoublepage


\renewcommand{\thefootnote}{\arabic{footnote}}
\setcounter{footnote}{0}



\pagestyle{plain} 
\setcounter{page}{1}
\pagenumbering{arabic}


%

\section{Introduction}
\label{sec:Introduction}

The measurement of \CP asymmetries in flavour-changing neutral-current processes
provides a crucial test of the Standard Model (SM). In particular, loop-mediated (penguin) decays of
$B$ mesons are sensitive probes for physics beyond the SM. Transitions between the quarks of the third and 
second generation ($b \to s$) or between the quarks of the third and first generation ($b \to d$)
are complementary since SM \CP violation is tiny in $b \to s$ transitions and an observation
of \CP violation would indicate physics beyond the SM. For $b \to d$
transitions the SM branching fraction is an order of magnitude smaller than $b \to s$ due to the
relative suppression of $|V_{td}|^2/|V_{ts}|^2$.
It is particularly useful to have experimental information on pairs
of channels related by $d \leftrightarrow s$ exchange symmetry to test that
the QCD contribution to the decay is independent of the initial \Bd or \Bs meson.

The \babar and \belle experiments have performed measurements of $b \to sq\overline{q}$ processes, such as
$\Bd \to \phi \KS$, $\Bd \to \eta^{\prime} \KS$ and
$\Bd \to f_0 \KS$~\cite{Abe:2003yt,Aubert:2005iy,Aubert:2005ja}, and of $b \to dq\overline{q}$ penguin diagrams, such as
$\Bd \to \KS \KS$ and $B^+ \to K^+ \KS$~\cite{Aubert:2006gm,Nakahama:2007dg}. These modes
contain pseudo-scalar or scalar mesons in their final state whereas $\Bzmeson \to VV^{\prime}$
decays, where $V$ and $V^{\prime}$ are light vector mesons,
provide a valuable additional source of information because the
angular distributions give insight into the physics of hadronic $B$ meson
decays and the interplay between the strong and weak interactions they involve. From the V$-$A
structure of the weak interaction and helicity conservation in the strong interaction,
the final state of these decays is expected to be highly longitudinally polarized.
This applies to both tree and penguin decays.
The \babar and \belle experiments have confirmed that longitudinal polarization dominates in
$b \to u$ tree processes such as $\Bd \to \rho^+ \rho^-$~\cite{Abe:2007ez,Aubert:2007nua},
$\Bp \to \rho^0 \rho^+$~\cite{Zhang:2003up,Aubert:2009it} and $\Bp \to \omega \rho^+$~\cite{Aubert:2009sx}.
However, measurements of the polarization in decays with both tree and penguin contributions, such as $\Bd \to \rho^0 \Kstarz$
and $\Bd \to \rho^- \Kstarp$~\cite{Lees:2011dq} and in $b \to s$ penguin
decays, \BdPhiKst~\cite{Chen:2005zv,BabarPhiKst2008}, \BsKstKst~\cite{Aaij:2011rf}
and \BsPhiPhi~\cite{Aaltonen:2011rs,LHCb-PAPER-2012-004,*Aaij:2013qha}, indicate a low value of the
longitudinal polarization fraction comparable with, or even smaller than, the transverse fraction.

The $\Bzmeson \to VV^{\prime}$ decays can be described by models based on
perturbative QCD, or QCD factorization and SU(3) flavour symmetries. Whilst some authors predict a
longitudinal polarization fraction $f_0\mathord{\sim}0.9$ for tree-dominated and $\mathord{\sim}0.75$
for penguin decays~\cite{Ali:1979al,*Suzuki:2002yk,Chen:2002pz}, other studies
have proposed different mechanisms such as penguin annihilation~\cite{Benm:2007rf,Cheng:2008gxa}
and QCD rescattering~\cite{Cheng:2004ru} to accommodate smaller longitudinal polarization fractions $\mathord{\sim}0.5$, although
the predictions have large uncertainties. A review on the topic of polarization in
$B$ decays can be found in Ref.~\cite{Beringer:1900zz}. 

There are only two other $\Bzmeson \to VV^{\prime}$ penguin modes that correspond to $b \to d$ loops. The 
first is the \BdKstKst decay. The \babar collaboration reported the discovery of this channel
with $6\,\sigma$ significance and a measurement of its branching fraction
$\BR(\BdKstKst)=(1.28\,{}^{+0.35}_{-0.30}\pm0.11) \times 10^{-6}$~\cite{PhysRevLett.100.081801}.
This is in tension with the results of the \belle collaboration that published
an upper limit of $\BR(\BdKstKst)<0.8 \times 10^{-6}$ at the $90\%$ confidence level~\cite{PhysRevD.81.071101}.
The \babar publication also reported a measurement of the longitudinal
polarization ${f_0 = 0.80^{+0.12}_{-0.13}}$~\cite{PhysRevLett.100.081801}, which is large compared to those from \BdPhiKst ($f_0 = 0.494 \pm 0.036$~\cite{BabarPhiKst2008}),
\BsPhiPhi ($f_0 = 0.365 \pm 0.025$~\cite{LHCb-PAPER-2012-004}) and \BsKstKst ($f_0 = 0.31 \pm 0.13$~\cite{Aaij:2011rf}).

\begin{figure}[!t]
\begin{center}
\includegraphics[scale=0.5]{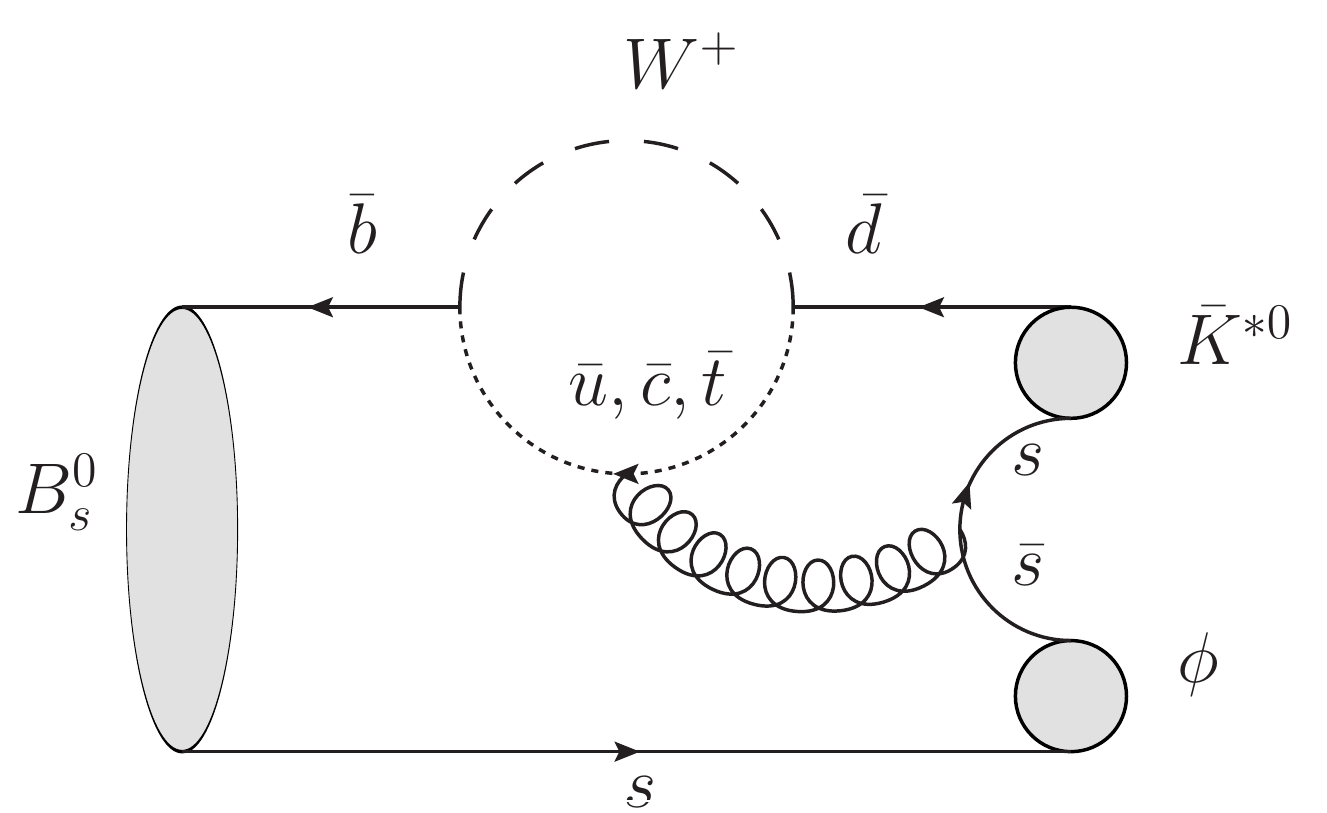}
\includegraphics[scale=0.5]{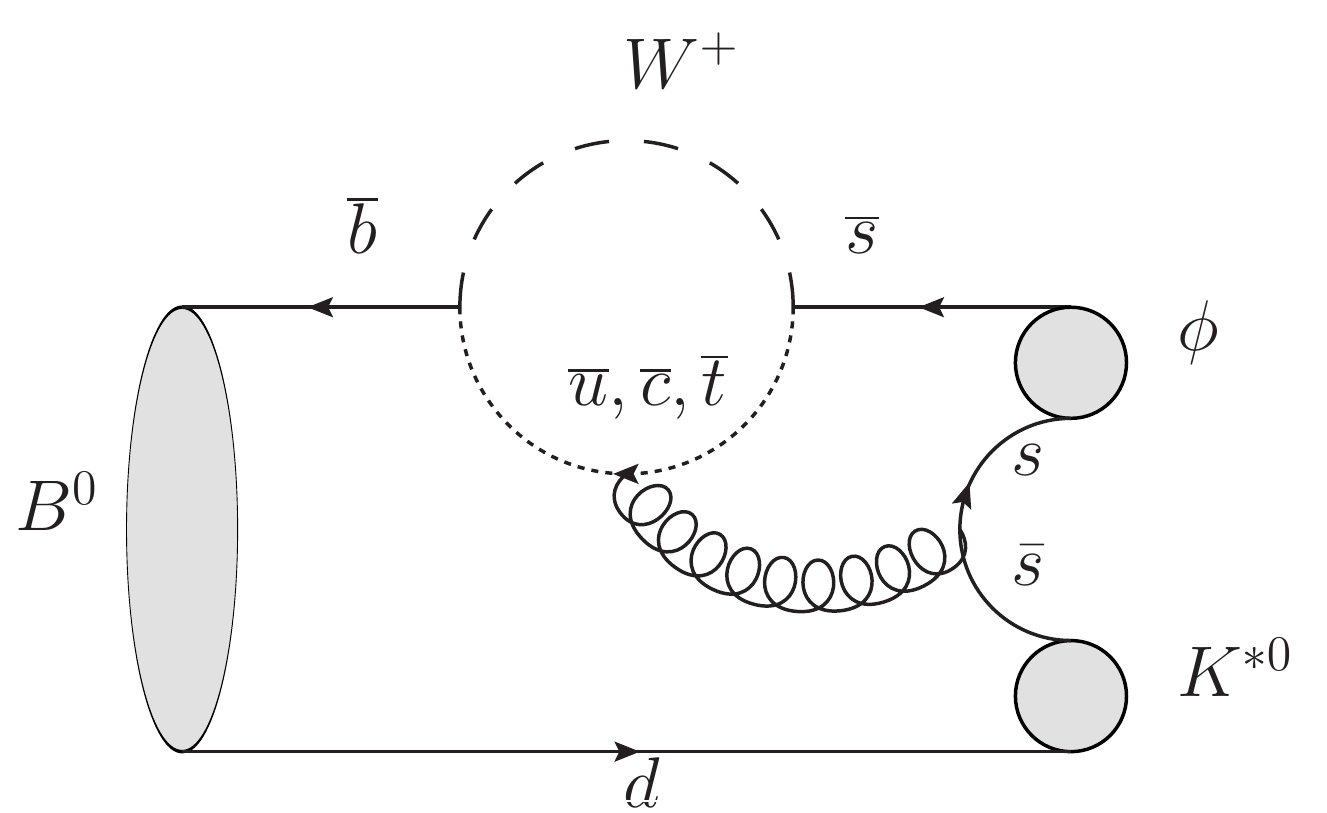}
\caption{\small Feynman diagrams for the \BsPhiKst and the \BdPhiKst decays.}
\label{fig:penguinBphiKst}
\end{center}
\end{figure}

The mode \BsPhiKst is the other $b \to d$ penguin decay into vector mesons that has not previously been
observed. This decay is closely linked to \BdPhiKst, differing in the spectator quark and
the final quark in the loop, as shown in Fig.~\ref{fig:penguinBphiKst}.\footnote{Both the decays \BsPhiKst and \BdPhiKst could also have
contributions from QCD singlet-penguin amplitudes~\cite{Benm:2007rf}.} From the aforementioned relation between $b \to s$ and $b \to d$ transitions,
their relative branching fractions should scale as $|V_{td}|^2/|V_{ts}|^2$ and their polarization fractions
are expected to be very similar. Moreover, since both decays share the same final state, except for charge conjugation,
\BdPhiKst is the ideal normalization channel for the determination of the
\BsPhiKst branching fraction. 
The \BsPhiKst decay is also related to \BdKstKst, since their loop diagrams only differ in the
spectator quark (${\squark \text{ instead of } \dquark}$), although it has been suggested that \swave interference effects
might break the SU(3) symmetry relating two channels~\cite{Gronau:1995hn}. Finally, it is also interesting to explore the relation of the \BsPhiKst
decay with the $\Bd \to \rho^0 \Kstarz$ mode since the penguin loop diagrams
of these modes are related by the $d \leftrightarrow s$ exchange. 
The $\Bd \to \rho^0 \Kstarz$ decay also has a $b \to u$ tree diagram, but it is expected that the penguin contribution is
dominant, since the branching fraction is comparable to that of the pure penguin \BdPhiKst decay.

The most stringent previous experimental limit on the \BsPhiKst branching fraction is
${\BF(\BsPhiKst)<1.0 \times 10^{-3}}$ at the $90\%$ confidence
level~\cite{Beringer:1900zz}, whereas calculations based on the
QCD factorization framework predict a value of ${(0.4\,{}^{+0.5}_{-0.3}) \times 10^{-6}}$~\cite{Benm:2007rf}
while in perturbative QCD a value of ${(0.65 \, {}^{+0.33}_{-0.23}) \times 10^{-6}}$~\cite{Ali:2007ff} is obtained.
The precise determination of the branching fraction tests these models and provides a probe for
physics beyond the SM.

The study of the angular distributions in the \BsPhiKst channel provides a measurement of its
polarization. In Ref.~\cite{Ali:2007ff}, a prediction of $f_0 = 0.712 \, {}^{+0.042}_{-0.048}$ is made for the longitudinal polarization fraction, using the
perturbative QCD approach, that can be compared to the experimental result.

In this paper the first observation of the \BsPhiKst decay, with $\phi \to \KK$ and $\Kstarzb \to \Kpi$, is reported and the
determination of its branching fraction and polarizations are presented.
The study is based on data collected by the \lhcb experiment at CERN from the $\sqrt{s} = 7\,$\tev proton-proton
collisions of \lhc beams. The dataset corresponds to an integrated luminosity of 1.0\,fb$^{-1}$.

\section{Detector and software}
\label{sec:Detector}

The \lhcb detector~\cite{Alves:2008zz} is a single-arm forward
spectrometer covering the \mbox{pseudorapidity} range $2<\eta <5$,
designed for the study of particles containing \bquark or \cquark
quarks. The detector includes a high precision tracking system
consisting of a silicon-strip vertex detector surrounding the $pp$
interaction region, a large-area silicon-strip detector located
upstream of a dipole magnet with a bending power of about
$4{\rm\,Tm}$, and three stations of silicon-strip detectors and straw
drift tubes placed downstream.
The combined tracking system provides a momentum measurement with
relative uncertainty that varies from 0.4\% at 5\gevc to 0.6\% at 100\gevc,
and impact parameter resolution of 20\mum for
tracks with high transverse momentum (\pt). Charged hadrons are identified
using two ring-imaging Cherenkov (RICH) detectors~\cite{Adinolfi:2012an}. Photon, electron and
hadron candidates are identified by a calorimeter system consisting of
scintillating-pad and preshower detectors, an electromagnetic
calorimeter and a hadronic calorimeter. Muons are identified by a
system composed of alternating layers of iron and multiwire
proportional chambers~\cite{Alves:2012ey}.

The trigger~\cite{Aaij:2012me} consists of a
hardware stage, based on information from the calorimeter and muon
systems, followed by a software stage, which applies a full event reconstruction.
The software trigger used in this analysis requires a two-, three- or four-track
secondary vertex with a high sum of the \pt of
the tracks and significant displacement from the primary $pp$
interaction vertices~(PVs). At least one track should have $\pt >
1.7\gevc$ and impact parameter \chisq~(\chisqIP) with respect to all
primary interactions greater than 16. The \chisqIP is defined as the
difference between the \chisq of a PV reconstructed with and
without the considered track. A multivariate algorithm~\cite{Gligorov:arXiv1210.6861} is used for
the identification of secondary vertices consistent with the decay
of a \bquark hadron.

In the simulation, $pp$ collisions are generated using
\pythia~6.4~\cite{Sjostrand:2006za} with a specific \lhcb
configuration~\cite{Belyaev:1307917}.  Decays of hadronic particles
are described by \evtgen~\cite{Lange:2001uf}, in which final state
radiation is generated using \photos~\cite{Golonka:2005pn}. The
interaction of the generated particles with the detector and its
response are implemented using the \geant
toolkit~\cite{Allison:2006ve, *Agostinelli:2002hh} as described in
Ref.~\cite{LHCb-PROC-2011-006}.

\section{Signal selection}
\label{sec:selection}

Signal \BsPhiKst candidates are formed from $\phi \to \KK$ and $\Kstarzb \to \Kpi$ decays.\footnote{Inclusion
of charge conjugated processes is implied in this work, unless otherwise stated.}
The pairs of charged particles in the $\phi \to \KK$ and the $\Kstarzb \to \Kpi$
candidates must combine to give invariant masses ${1012.5 < M(\KK) < 1026.5\mevcc}$ and ${746 < M(\Kpi)< 1046\mevcc}$,
consistent with the known $\phi$ and \Kstarzb masses~\cite{Beringer:1900zz}.
Each of the four tracks is required to have $\pt > 500\mevc$
and \chisqIP$>9$.

Kaons and pions are distinguished by use of
a log-likelihood algorithm that combines information from the RICH detectors
and other properties of the event~\cite{Adinolfi:2012an}.
The final state particles are identified by requiring that the
difference in log-likelihoods of the kaon and pion mass hypotheses is \dllkpi$>2$
for each kaon candidate and $<0$ for the pion candidate.
In addition, the difference in log-likelihoods of the proton and
kaon hypotheses, \dllpk, is required to be $<0$ for the kaon from the
\Kstarzb decay. This suppresses background
from \Lb decays. This requirement is not necessary for the kaons from the $\phi$
candidate owing to the narrow \KK invariant mass window.

The \Kpi pair that forms the \Kstarzb candidate is required to originate from
a common vertex with a $\chisq$ per number of degrees of freedom ($\chisq/{\rm ndf}$) $<9$, and to have a positive
cosine of the angle between its momentum and the reconstructed \Bzmeson candidate flight direction,
calculated with the \Bzmeson decay vertex and the best matching primary vertex.
The \Kpi combination is also required to have $\pt > 900\mevc$.
The same conditions are imposed on the $\phi$ candidate.

The \Bzmeson candidates are also required to fulfil some minimal selection
criteria: the $\phi$ and \Kstarzb candidates must form a vertex with $\chisq/{\rm ndf}<15$;
the distance of closest approach between their trajectories must be less than $0.3\,\mm$;
and they must combine to give an invariant mass within ${4866 < M(\KK\Kpi) < 5866\mevcc}$.

In addition, a geometrical-likelihood based selection (GL)~\cite{Karlen:1998zz,MartinezSantos:1264603}
is implemented using as input variables properties of the \Bzmeson meson candidate. These are
\begin{itemize}

\item the \Bzmeson candidate impact parameter (\IP) with respect to the closest primary vertex;

\item the decay time of the \Bzmeson candidate;

\item the \pt of the \Bzmeson candidate;

\item the minimum \chisqIP of the four tracks with respect to all
primary vertices in the event; and

\item the distance of closest approach between the \Kstarzb and $\phi$ candidates' trajectories
reconstructed from their respective daughter tracks.

\end{itemize}

The GL is trained to optimize its discrimination power using representative
signal and background samples. For the signal a set of \BsPhiKst simulated events is
used. For the background a sample of events where, in addition to the signal selections,
other than those on the masses, requirements of $999.5<M(\KK) < 1012.5 \mevcc$ or ${1026.5<M(\KK) < 1039.5 \mevcc}$
for the $\phi$ candidate and ${M(\KK\Kpi)>5413\mevcc}$ for the four-body mass are applied. The selection of
only the high-mass \Bzmeson sideband is motivated by the nature of the background in
that region, which is purely combinatorial, whereas the low-mass sideband contains
partially reconstructed $B$ meson decays that have topological similarities to the signal.

\section{Suppression of background from other {\boldmath \bquark}-hadron decays}
\label{subsec::addsel}

A small background from \BsPhiPhi decays, where one of the kaons from the $\phi$ is
misidentified as a pion, is found to contaminate the signal.
Candidate \BsPhiKst decays are therefore required to be outside of the window defined by
${1012.5 < M(\KK) < 1026.5\mevcc}$ and ${5324 < M(\KK\KK) < 5424 \mevcc}$
in the \KK and $\KK\KK$ invariant masses when the mass
hypothesis for the sole pion of the decay is switched into a kaon.
In simulated events this selection removes $0.12 \%$ of
the \BsPhiKst signal decays and does not affect the \BdPhiKst decay mode.
Other possible reflections, such as \BsKstKst decays, are found to be negligible.

In order to remove background from $\Bs \to D_s^{\mp}(\phi\pi^{\mp}) K^{\pm}$ decays when the $\pi^{\mp}$ and the
$K^{\pm}$ mesons form a $\brabar{K}^{*0}$ candidate, events with the invariant mass of the 
$\KK\pi^{\mp}$ system within ${1953.5 < M(\KK\pi^{\mp}) < 1983.5 \mevcc}$, consistent
with the known \Ds mass~\cite{Beringer:1900zz}, are excluded.

Background from $b$-hadron decays containing a misidentified proton has also been considered.
For candidate \BsPhiKst decays, the kaon with the largest \dllpk is assigned the proton mass and the four-body invariant mass recomputed.
The largest potential background contribution arises from $\Lbbar \to \KK \antiproton \pi^+$ where the antiproton is
misidentified as the kaon originating from the \Kstarzb meson, and $\Lb \to \KK K^-\proton$, where the proton
is misidentified as the pion originating from the \Kstarzb meson.
Simulation shows that these decays produce wide four-body mass distributions
which peak around $5450\mevcc$ and $5500\mevcc$, respectively.
This background contribution is considered in the fit model discussed below.
Other \Bzmeson  decay modes containing a $\L \to \proton \pi^-$ decay or  
background from $\Lc \to \proton\Kpi$ decays are found to be negligible.

\section{Fit to the four-body mass spectrum}
\label{sec:mass-spectrum}

The sample of $1277$ candidates, selected as described in
Sections~\ref{sec:selection} and~\ref{subsec::addsel}, contains many \BdPhiKst
decays whereas only a small contribution from \BsPhiKst decays is anticipated. Both signals
are parametrized with identical shapes, differing only in the mass shift of $87.13\mevcc$
between the \Bd and \Bs mesons~\cite{Beringer:1900zz} which is fixed in the fit.
The signal shapes are described by the sum of Crystal Ball (CB)~\cite{Skwarnicki:1986xj}
and \gaussian functions that share a common mean.  The CB function, which contains most
of the signal, is a combination of a \gaussian function with a power law tail, accounting
for the intrinsic detector resolution and the radiative tail toward low masses, respectively.
The \gaussian shape describes events reconstructed with worse mass resolution, which produce
a contamination of \BdPhiKst decays in the region of the \BsPhiKst signal peak.
The dependence between the \gaussian and CB resolutions, $\sigma_{{\rm G}}$ and $\sigma_{{\rm CB}}$,
respectively, is found to be 
\begin{equation}
\label{eq:sigmas}
\sigma_{{\rm G}} = \sqrt{\sigma_{{\rm CB}}^2 + (24.74\mevcc)^2},
\end{equation}
from a data sample of $25 \times 10^{3}$ \BdToJPsiKst decays.
This channel is topologically very similar to the signal and is almost background free.
The fit to this sample also provides the power law exponent of the CB function tail, which is subsequently fixed
in the $\Bs \to (\KK)(\Kpi)$ and $\Bd \to (\KK)(\Kpii)$ mass models. The parameter that governs the transition from the Gaussian shape to the
power law function in the CB function is unrestrained in the fit. 
The other unrestrained fit parameters include: the central $B$ meson mass,
the width of the CB function, the fractional yield contained in the \gaussian function and the total signal yield.

In addition to the \Bd and \Bs signal shapes, three more components are included.
The first accounts for partially reconstructed $B$ meson decays into $\phi$ and $K$ or $K^*$
excited states where a pion has been lost.
This is described by a convolution of the ARGUS shape~\cite{Albrecht:1990cs} with a \gaussian distribution. 
The second contribution is due to $\Lb \to \KK K^-\proton$  and $\Lbbar \to \KK \antiproton \pi^+$ decays and is modelled with a histogram
obtained from simplified simulations.
The third contribution is an exponential function to account for combinatorial background.

\begin{figure}[t]
\centering
\includegraphics[width=0.6\textwidth]{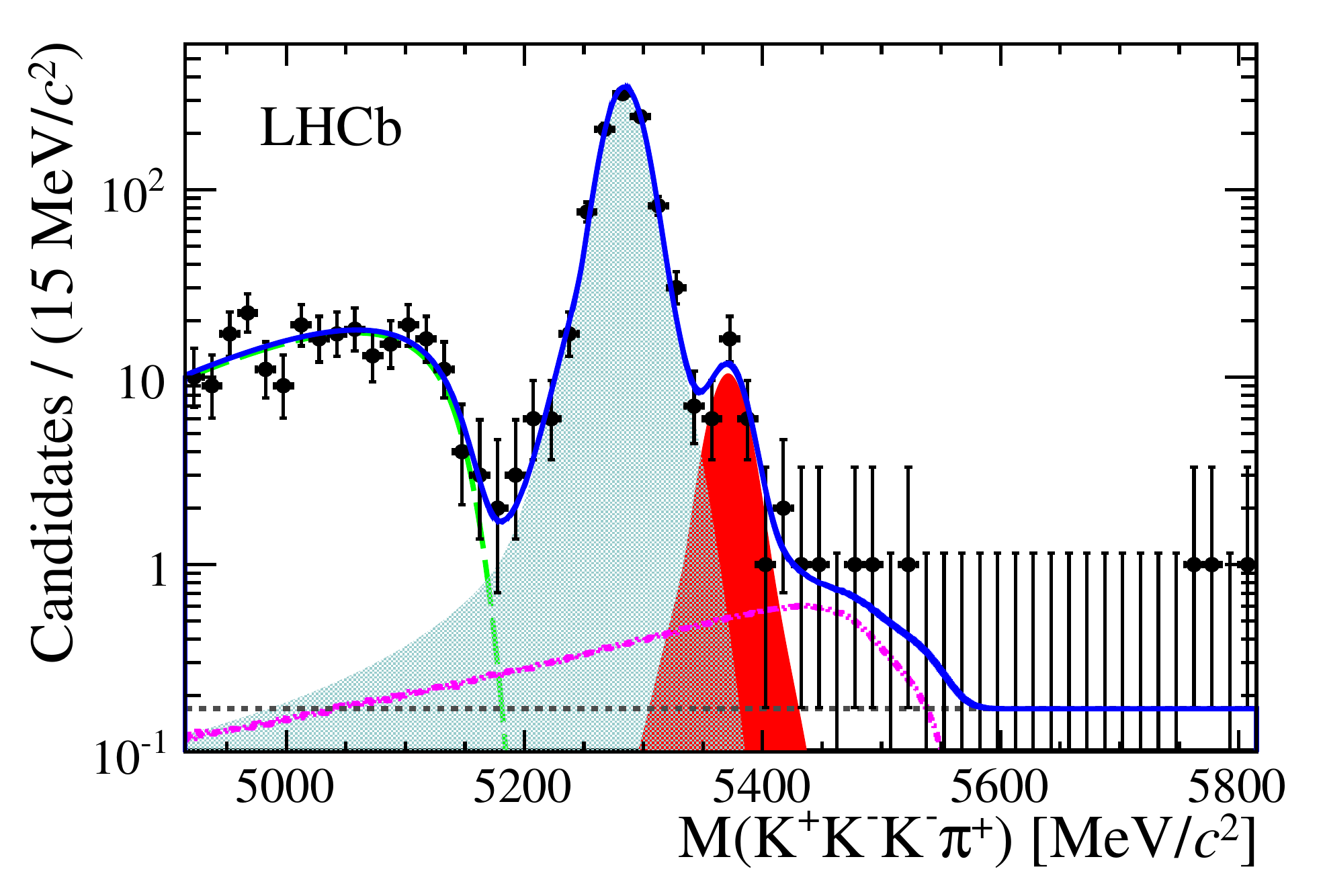}
\caption{\small Four-body $\KK\Kpi$ invariant mass distribution. The points show the data, the blue solid line shows the overall fit, the
solid dark red shaded region is the \BsPhiKst signal, the light blue shaded region corresponds to the \BdPhiKst signal, the
grey dotted line is the combinatorial background and the green dashed line and magenta dashed-dotted
lines are the  partially reconstructed and misidentified \Lb backgrounds.}
\label{fig::fig_final}
\end{figure}

The data passing the selection criteria are fitted using an extended unbinned maximum likelihood fit.
The invariant mass distribution of the candidates, together with the fit contribution,
is shown in Fig.~\ref{fig::fig_final}. The yields of ${\Bs \to (\KK)(\Kpi)}$
and ${\Bd \to (\KK)(\Kpii)}$ decays are $30 \pm 6$ and $1 000 \pm 32$, respectively.
The fit model is validated with $10,000$ pseudo-experiments, generated with simplified simulations,
which show that the signal yields are unbiased. Table~\ref{tab::fit_results} summarizes the signal
and background contributions resulting from the fit. A likelihood ratio test is employed to assess
the statistical significance of the ${\Bs \to (\KK)(\Kpi)}$ signal yield.
This is performed using $\sqrt{2{\rm ln}(\mathcal{L}_{\rm s+b}/\mathcal{L}_{\rm b})}$, where $\mathcal{L}_{\rm s+b}$ and $\mathcal{L}_{\rm b}$
are the maximum values of the likelihoods for the signal-plus-background
and background-only hypotheses, respectively.\footnote{The applicability of this method has
been verified from the parabolic behaviour of the ${\Bs \to (\KK)(\Kpi)}$ signal yield
 profile of $-2\ln \mathcal{L}_{\rm s+b}$ about its minimum.} This calculation results in
$6.3\,\sigma$ significance for the ${\Bs \to (\KK)(\Kpi)}$ signal. The fit gives $\sigma_{\rm CB} = 15.0 \pm 1.1\mevcc$ for the
invariant mass resolution.
Integration in a $\pm 30\mevcc$ mass window yields $26.4 \pm 5.7$ signal
candidates and $8.2 \pm 1.3$ background events, composed of $5.4 \pm 0.2$ from
${\Bd \to (\KK)(\Kpii)}$, $2.1 \pm 1.3$ from \Lb and $0.7 \pm 0.4$ from combinatorial contributions.

In order to explore systematic effects in the signal yield originating in the fit model two effects were considered.   
First, the amount of ${\Bd \to (\KK)(\Kpii)}$ events under the ${\Bs \to (\KK)(\Kpi)}$ signal is governed
by the $24.74 \mevcc$ factor in Eq.~\ref{eq:sigmas}. Similarly, the contamination of misidentified \Lb decays
under the signal is controlled by a tail that is parametrized. An extended likelihood is built by multiplying the original likelihood
function by \gaussian distributions of these two nuissance parameters with standard deviations
of $20\%$ of their  nominal values at which they are centered. The corresponding systematic uncertainty in the
signal yield is obtained by performing a fit that maximizes this modified likelihood.
The systematic contribution is calculated subtracting the statistical uncertainty in quadrature and found to be $\pm 1.2$ events.
Including this uncertainty results in a significance of $6.2\sigma$. Effects of other systematic uncertainties, discussed
in Sect.~\ref{sec:Kpolarization}, have negiglible impact in the signal significance.

\begin{table}
\centering
\caption{\small Results of the fit to the sample of selected candidates.}
\begin{tabular}{cc}
\hline \hline
Contribution & Yield \\
\hline 
\vspace{-0.4cm}
& \\
  \BsPhiKst & $\:\: 30 \pm 6$ \\
  \BdPhiKst & $1000 \pm  32$ \\
  Partially reconstructed background &  $\:\: 218 \pm 15$ \\
  \Lb background & $\:\: 13 \pm 8$ \\
  Combinatorial background & $\:\: 10 \pm 6$ \\
\hline
\hline
\end{tabular}
\label{tab::fit_results}
\end{table}

\section{Determination of the S-wave contribution}
\label{sec:Purity}

The $\Bs \to (\KK)(\Kpi)$ signal is expected to be
mainly due to \BsPhiKst decays, although there are possible non-resonant
contributions and \KK and \Kpi pairs from other resonances.
To estimate the \swave contributions, it is assumed that the effect
is the same for \BdPhiKst and \BsPhiKst decays, therefore
allowing the larger sample of \BdPhiKst decays to be used. 
The effect of this assumption is considered as a source of systematic uncertainty
in Sect.~\ref{sec:syst}.

The \KK invariant mass distribution for $\phi$ candidates within a $\pm 30\mevcc$ window
of the known \Bd mass is described by a relativistic spin-1 Breit-Wigner distribution convolved with a \gaussian
shape to account for the effect of resolution. A linear term is added to describe the \swave contribution.
The purity resulting from this fit is $0.95 \pm 0.01$ in a $\pm7\mevcc$ window around the known $\phi$ mass.

The \Kpii pairs are parametrized by the incoherent sum of a relativistic spin-1 Breit-Wigner amplitude and
a shape that describes non-resonant and $\Kstarz(1430)$ \swave contributions
introduced by the LASS experiment~\cite{BabarPhiKst2008,LASS:1988}.
The fraction of events from \Kstarz decays within a $\pm150\mevcc$ window around the \Kstarz mass
results in a purity of $0.89 \pm 0.02$.
When combining the \KK and \Kpii contributions, the total $\phi K^{*0}$ purity is found to be $0.84 \pm 0.02$.
This purity can be translated into a p-value, quantifying the probability that the entire
${\Bs \to (\KK)(\Kpi)}$ signal is due to decays other than $\phi \Kstarzb$. After combining with
the ${\Bs \to (\KK)(\Kpi)}$  significance the \BsPhiKst is observed with $6.1 \, \sigma$ significance. 

\begin{figure}[t]
  \centering
    \includegraphics[width=0.4\textwidth]{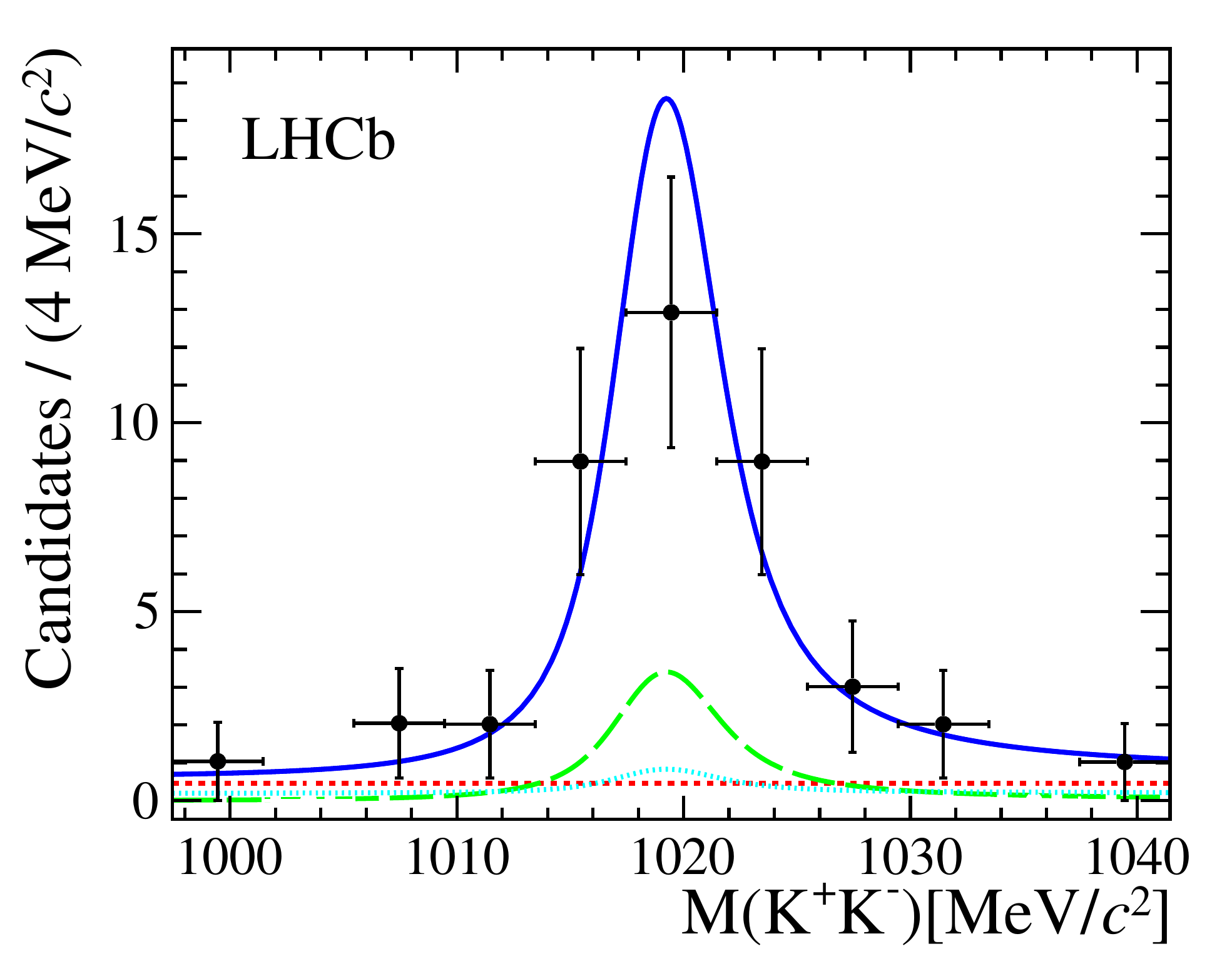}
    \includegraphics[width=0.4\textwidth]{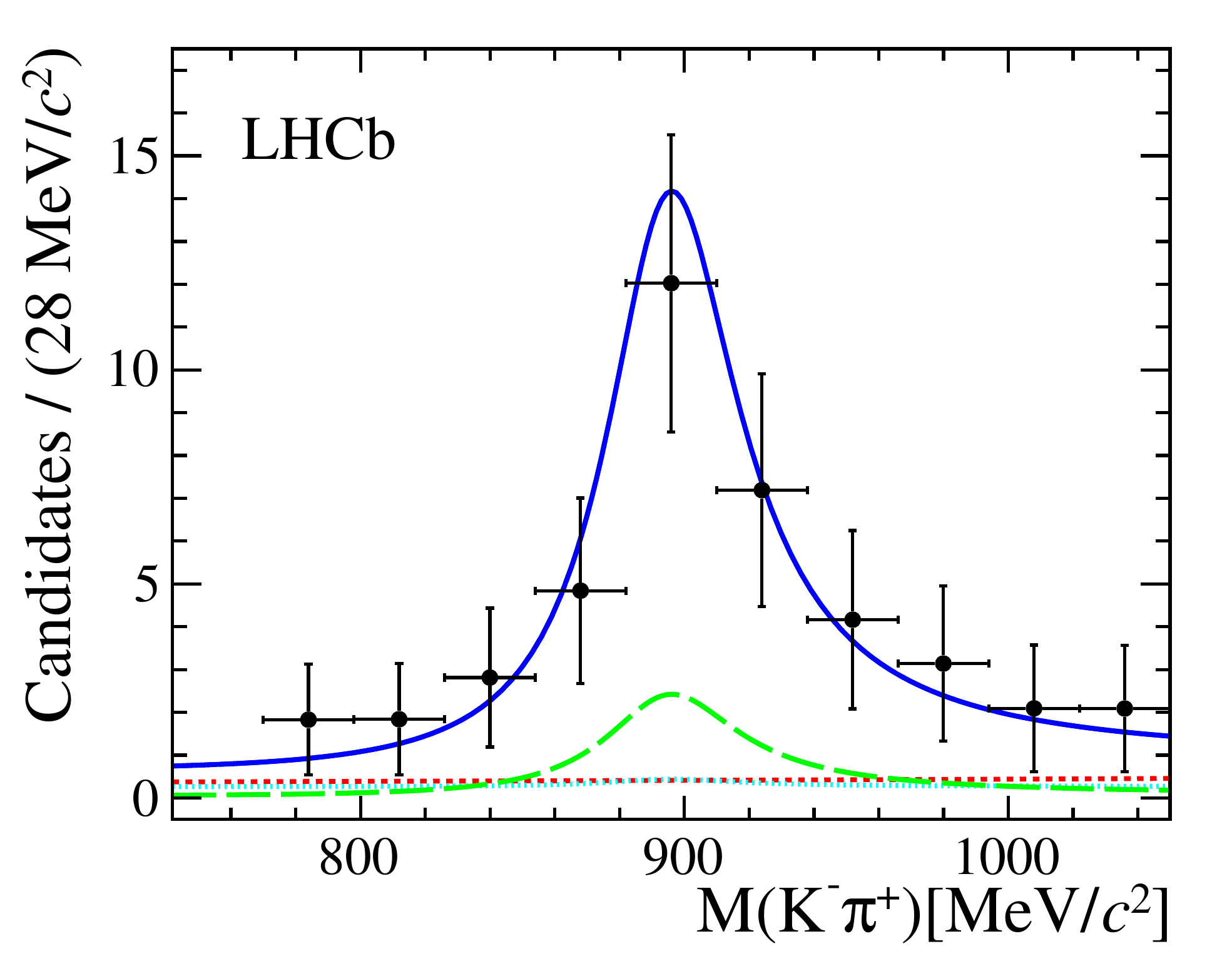}
    \includegraphics[width=0.4\textwidth]{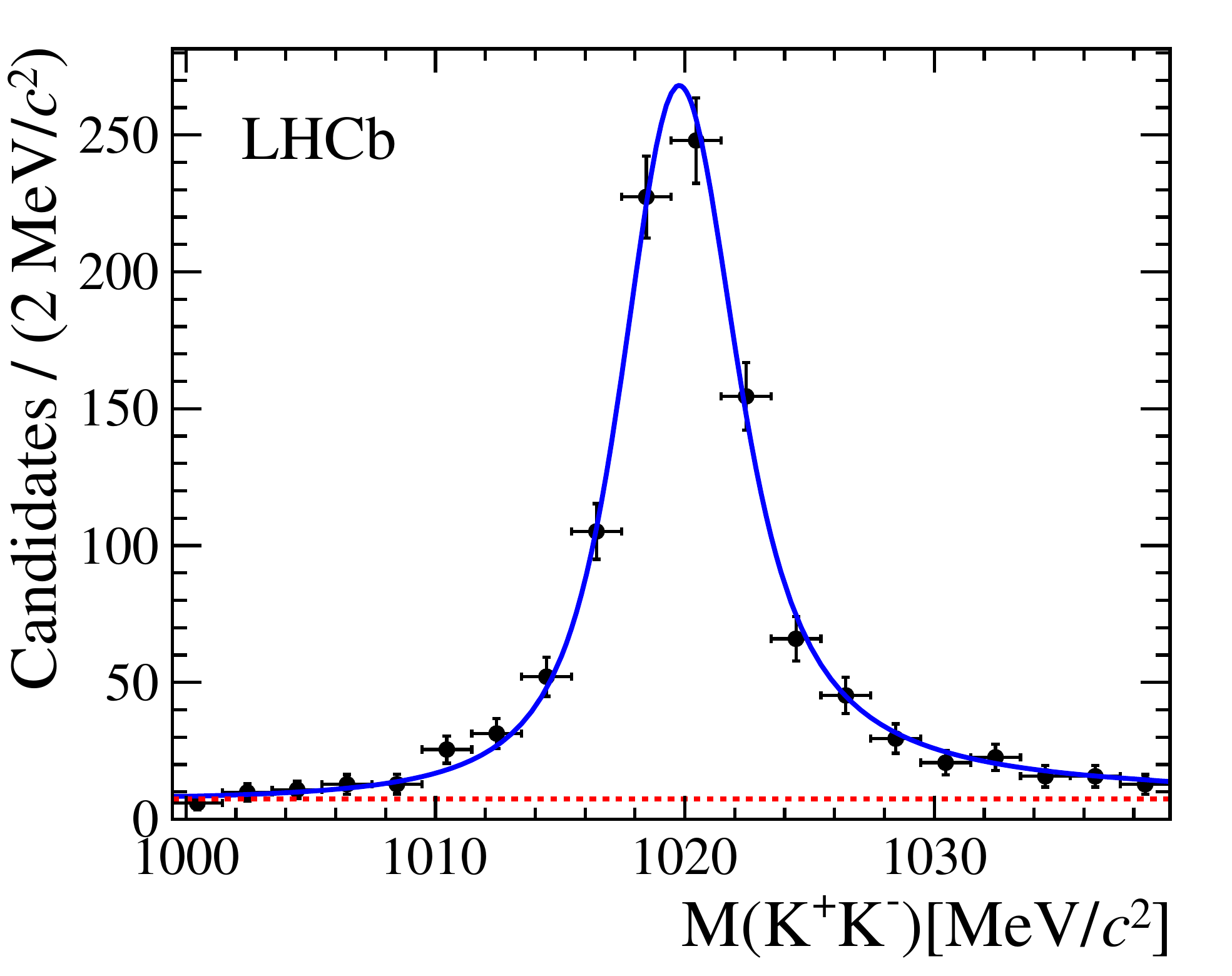}
    \includegraphics[width=0.4\textwidth]{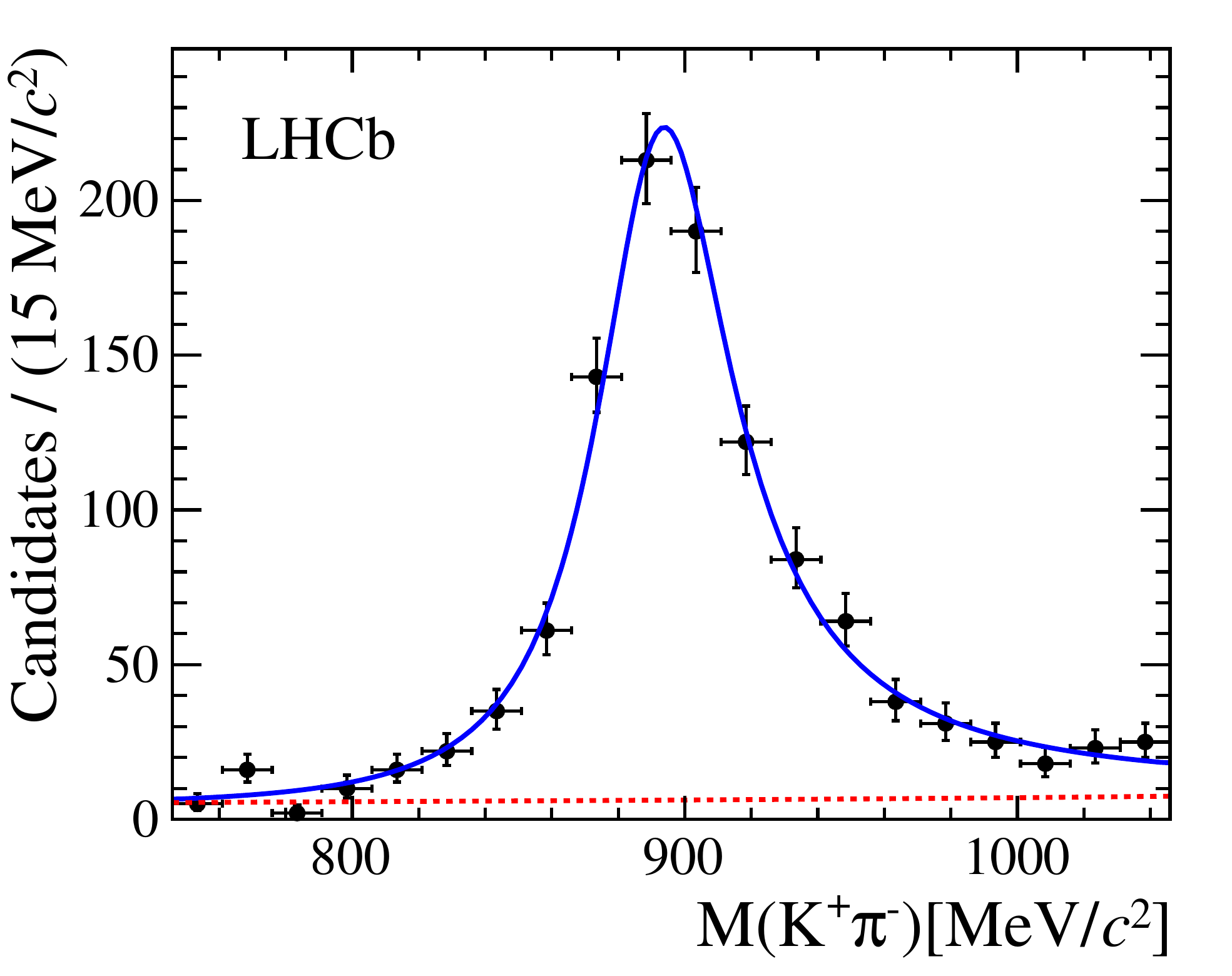}
\label{fig:kkkpimasses}
  \caption{\small Invariant mass distributions for (left) \KK and (right) $K^{\mp}\pi^{\pm}$ pairs
in a $\pm 30\mevcc$ window around the (top) \Bs and (bottom) \Bd mass. The solid blue line is the overall fit, 
the green dashed line corresponds to \Bd cross-feed into the \Bs mass window, 
the red dotted line is the \swave contribution and the light blue is the combinatorial background.}
\end{figure}

\section{Determination of the {\boldmath\BsPhiKst} branching fraction}
\label{sec:Branching}

The branching fraction is calculated with the \BdPhiKst channel as
normalization.
Both decays pass the same selection and share almost identical topologies.
However, since the two decay channels can have different polarizations, their
angular distributions may differ which would cause a difference in their detection efficiencies.
A factor 
\begin{equation}
\lambda_{f_0}= \frac{\epsilon^{\BdPhiKst}}{\epsilon^{\BsPhiKst}} = \frac{1-0.29
f_0^{\BdPhiKst}}{1-0.29 f_0^{\BsPhiKst}}
\nonumber
\end{equation}
\noindent
is calculated, where $\epsilon^{\BdPhiKst}$ and
$\epsilon^{\BsPhiKst}$ are
the efficiencies for the \BdPhiKst and \BsPhiKst decays reconstruction,
$f_0^{\BdPhiKst}$ and $f_0^{\BsPhiKst}$ their longitudinal polarization
fractions, determined in Sect.~\ref{sec:Kpolarization} for the \BsPhiKst mode, and the
factor 0.29 is obtained from simulation.

The value of $\BF(\BsPhiKst)$ is computed from
\begin{equation}
\BF(\BsPhiKst) =  \lambda_{f_0} \times \frac{f_d}{f_s} \times
\BF(\BdPhiKst) \times \frac{N_{\BsPhiKst}}{N_{\BdPhiKst}},
\label{BRformulaBsBd}
\end{equation}

\noindent where $N_{\BsPhiKst}$ and $N_{\BdPhiKst}$ are the numbers of \Bs and
\Bd decays, respectively, and $f_d/f_s = 3.75 \pm 0.29 $~\cite{LHCbfsfd} 
is the ratio of hadronization factors needed to account for the different
production rates of \Bd and \Bs mesons.
With the values given in Table~\ref{table_BRinput}, the result,
\[
\BF(\BsPhiKst) = (1.10 \pm 0.24) \times 10^{-6},
\]

\noindent is obtained, where only the statistical uncertainty is shown.

\begin{table}
\begin{center}
\caption{\small Input values for the branching fraction computation.}
\begin{tabular}{cc}
  \hline
  \hline
   Parameter & Value \\
\hline
   $\lambda_{f_0}$          &  $ 1.01 \pm 0.06$ \\
   $N_{\BdPhiKst}$          &  $1000 \pm  32\:\:\:\:$ \\
   $N_{\BsPhiKst}$          &  $30 \pm 6\:\:$ \\
   ${\cal B}(\BdPhiKst)$    &  $(9.8 \pm 0.6) \times 10^{-6}$~\cite{Beringer:1900zz} \\
  \hline
  \hline
\end{tabular}
\label{table_BRinput}
\end{center}
\end{table}

As a cross-check, a different decay mode, \BdToJPsiKst, with $J/\psi \to \mu^+ \mu^-$, has been
used as a normalization channel. Special requirements were imposed to harmonize the selection of this reference
with that for the signal.  The obtained result is fully compatible with the \BdPhiKst based value.

\section{Systematic uncertainties on the branching fraction}
\label{sec:syst}

Four main sources of systematic effects in the determination of the
branching fraction are identified: the fit model, the dependence of the acceptance on the longitudinal polarization,
the purity of the signal and the uncertainty in the relative efficiency of \Bs and \Bd detection.

Alternatives to the fit model discussed in Sect.~\ref{sec:mass-spectrum}
give an uncertainty of $\pm 1.2$ in the signal yield. This results in a
relative systematic uncertainty of $\pm 0.04$ on the branching fraction.

The systematic uncertainty in the acceptance correction
factor $\lambda_{f_0}$ originates from the uncertainties of the longitudinal polarization
fractions, $f_0$, in the  \BsPhiKst and \BdPhiKst channels and is found to be $\pm 0.06$.

As described in Sect.~\ref{sec:Purity} an \swave contribution of $0.16 \pm 0.02$ was found in the
\KK and \Kpi mass windows of the \BdPhiKst candidates.
The uncertainty caused by the assumption that this fraction is the same in \Bd
and \Bs decays is estimated to be $50\%$ of the
\swave contribution. This results in a $\pm 0.08$ contribution to the systematic uncertainty.
This uncertainty also accounts
for uncanceled interference terms between the
\Kstarz, the $\phi$ and their corresponding {\swave}s. These contributions are linear in the
sine or cosine of polarization angles~\cite{BabarPhiKst2008} and cancel after integration.
The dependence of the
acceptance on the angles violates this cancellation contributing
$\pm 0.04$ to the total $\pm 0.08$ \swave uncertainty.

The \BsPhiKst and \BdPhiKst final states are very similar and a detector
acceptance efficiency ratio $\sim 1$ is expected.
However, small effects, such as the mass shift $M(\Bs)-M(\Bd)$, 
translate into slightly different \pt distributions for the daughter particles.
This results in an efficiency ratio of $1.005$, as determined from simulation.
The deviation of $\pm 0.005$ from unity is taken as a systematic uncertainty that is propagated to
the branching fraction.

Finally, the uncertainty in the knowledge of the \BdPhiKst decay branching fraction of $\pm 0.6 \times 10^{-6}$ 
is also accounted for and results in a relative uncertainty of 0.06 in the \BsPhiKst decay branching fraction.

A summary of the systematic uncertainties is shown in Table~\ref{table_systsum}.
The final result for the \BsPhiKst decay branching fraction is
\[
\BF(\BsPhiKst) = \left(1.10 \pm 0.24\,\stat \pm 0.14\,\syst \pm 0.08\left(\frac{f_d}{f_s}\right)\right) \times 10^{-6},
\]

\noindent which corresponds to a ratio with the \BdPhiKst decay branching fraction of:
\[
\frac{\BF(\BsPhiKst)}{\BF(\BdPhiKst)} = 0.113 \pm 0.024\,\stat \pm 0.013\,\syst \pm 0.009\left(\frac{f_d}{f_s}\right).
\]

\begin{table}[t]
\begin{center}
\caption{\small Sources of systematic uncertainty in the branching fraction measurement.
The total uncertainty is the addition in quadrature of the individual sources.}
\begin{tabular}{cc}
  \hline
  \hline
   Source  & Relative uncertainty in \BR \\
  \hline
   Fit model  & $0.04\phantom{0}$ \\
   $f_0$  & $0.06\phantom{0}$ \\
   Purity  & $0.08\phantom{0}$ \\
   Acceptance  & $0.005$ \\
   \BF(\BdPhiKst) & $0.06\phantom{0}$ \\
  \hline
   Total & $0.12\phantom{0}$ \\
  \hline
  \hline
\end{tabular}
\label{table_systsum}
\end{center}
\end{table}

\section{Polarization analysis}
\label{sec:Kpolarization}

\begin{figure}[t]
  \centering
    \includegraphics[width=0.5\textwidth]{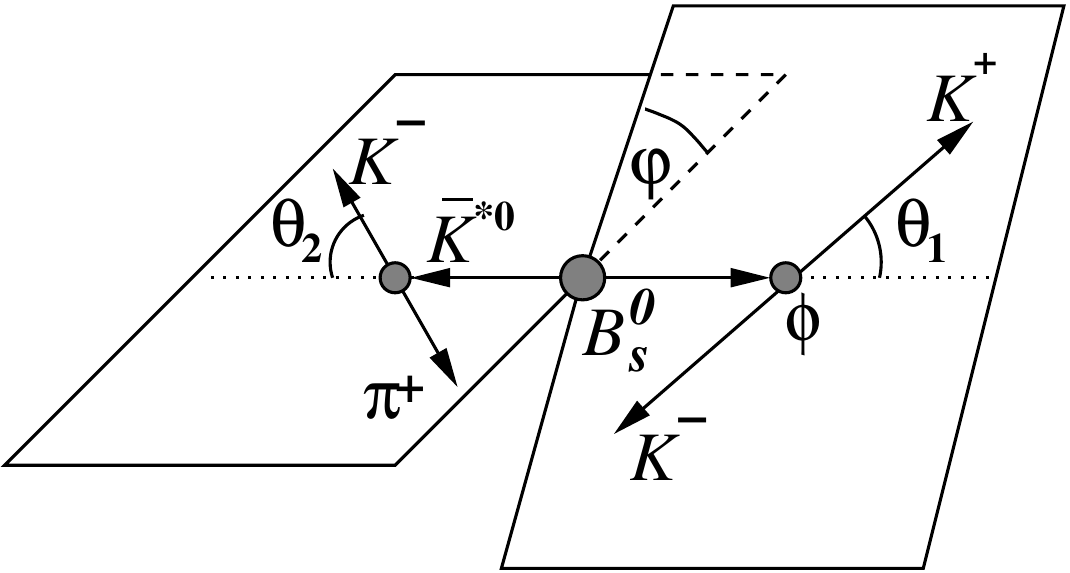}
\label{fig:anglesConv}
  \caption{\small Definition of the angles in \BsPhiKst decays where $\theta_1$ ($\theta_2$) is the
$K^+$ ($K^-$) emission angle with respect to the direction opposite to the \Bs
meson in the $\phi$ (\Kstarzb) rest frame and $\varphi$ is the angle between
the \Kstarzb and $\phi$ decay planes in the \Bs rest frame.}
\end{figure}

The $\BsPhiKst \to (\KK)(\Kpi)$ decay proceeds via two intermediate
spin-1 particles. The angular distribution of the decay
is described by three transversity amplitudes $A_0$, $A_{\parallel}$ and $A_{\perp}$~\cite{Dighe:1995pd}.
These can be obtained from the distribution of the decay products in three
angles $\theta_1$, $\theta_2$ and $\varphi$, defined in the helicity frame. The convention for the angles is shown in Fig.~\ref{fig:anglesConv}.
A flavour-averaged and time-integrated polarization analysis is performed
assuming that the \CP-violating phase is zero and that an equal amount of \Bs
and \Bsb mesons are produced.
Under these assumptions, the decay
rate dependence on the polarization angles can be written as
%
\begin{eqnarray}
\label{formula:fullAngularEq}
\frac{{\rm d}^3\Gamma}{{\rm d}{\rm cos}\theta_1 \, {\rm d}{\rm cos}\theta_2 \,
{\rm d}\varphi} &\propto& |A_0|^2\cos^2\theta_1\cos^2\theta_2 + |A_{\parallel}|^2 \frac{1}{2}\sin^2\theta_1\sin^2\theta_2\cos^2 \varphi \\
&+& |A_{\perp}|^2\frac{1}{2}\sin^2\theta_1\sin^2\theta_2\sin^2\varphi +
|A_0||A_{\parallel}|\cos\delta_{\parallel}
\frac{1}{2\sqrt{2}}\sin 2\theta_1\sin 2\theta_2\cos\varphi.
\nonumber
\end{eqnarray}

\noindent 
Additional terms accounting for the \swave and
interference contributions, as in Ref.~\cite{BabarPhiKst2008}, are also considered.
These terms are set to the values obtained for the \BdPhiKst sample.

The polarization fractions are defined from the amplitudes as:
$f_j=|A_j|^2/(|A_0|^2+|A_{\parallel}|^2+|A_{\perp}|^2)$ (with $j=0,\parallel,\perp$).
In addition to the polarization fractions the cosine of the phase difference between $A_0$ and $A_{\parallel}$, $\cos\delta_\parallel$, is accessible in this study.

The determination of the angular amplitudes depends on the spectrometer acceptance
as a function of the polarization angles $\theta_1$ and $\theta_2$. The acceptance was found not to depend on $\varphi$.
A parametrization of the acceptance as a function of $\theta_1$ and $\theta_2$
is calculated using simulated data and is used to correct the differential decay rate by
scaling Eq.~\ref{formula:fullAngularEq}.
Additionally, a small correction for discrepancies
in the $p_{\rm T}$ spectrum and the trigger selection of the $B$ mesons between simulation and data is introduced.

The data in a $\pm 30\mevcc$ window around the \Bs mass
are fitted to the final angular distribution. The fit accounts for
two additional ingredients: the tail of the \BdPhiKst decays, that are polarized with
a longitudinal polarization fraction of $f_0 = 0.494$~\cite{BabarPhiKst2008}, and 
the combinatorial background, parametrized from the distributions of events in
the high-mass $B$ sideband $5450 < M(\KK\Kpi) < 5840\mevcc$ after relaxing the selection
requirements. The latter accounts for both the combinatorial and misidentified \Lb backgrounds.

The systematic uncertainties in the determination of the angular
parameters are calculated modifying the analysis and computing the difference with the nominal result.
Three elements are considered.
\begin{itemize}
\item The uncertainty in the \swave fraction. This is computed modifying the 
\swave contribution by $50\%$ of its value. This covers within $2\,\sigma$ an \swave fraction 
from 0 to $30\%$, consistent with that typically found in decays of $B$ mesons
to final states  containing a \Kstarz meson.

\item The spectrometer acceptance. This contribution is calculated comparing the results considering
or neglecting the above-mentioned $p_{\rm T}$ and trigger corrections to the acceptance.
\item The combinatorial background. The background model derived from the $B$ mass
sideband is replaced by a uniform angular distribution. 
\end{itemize}

\begin{figure}[t!]
  \centering
    \includegraphics[scale=0.35]{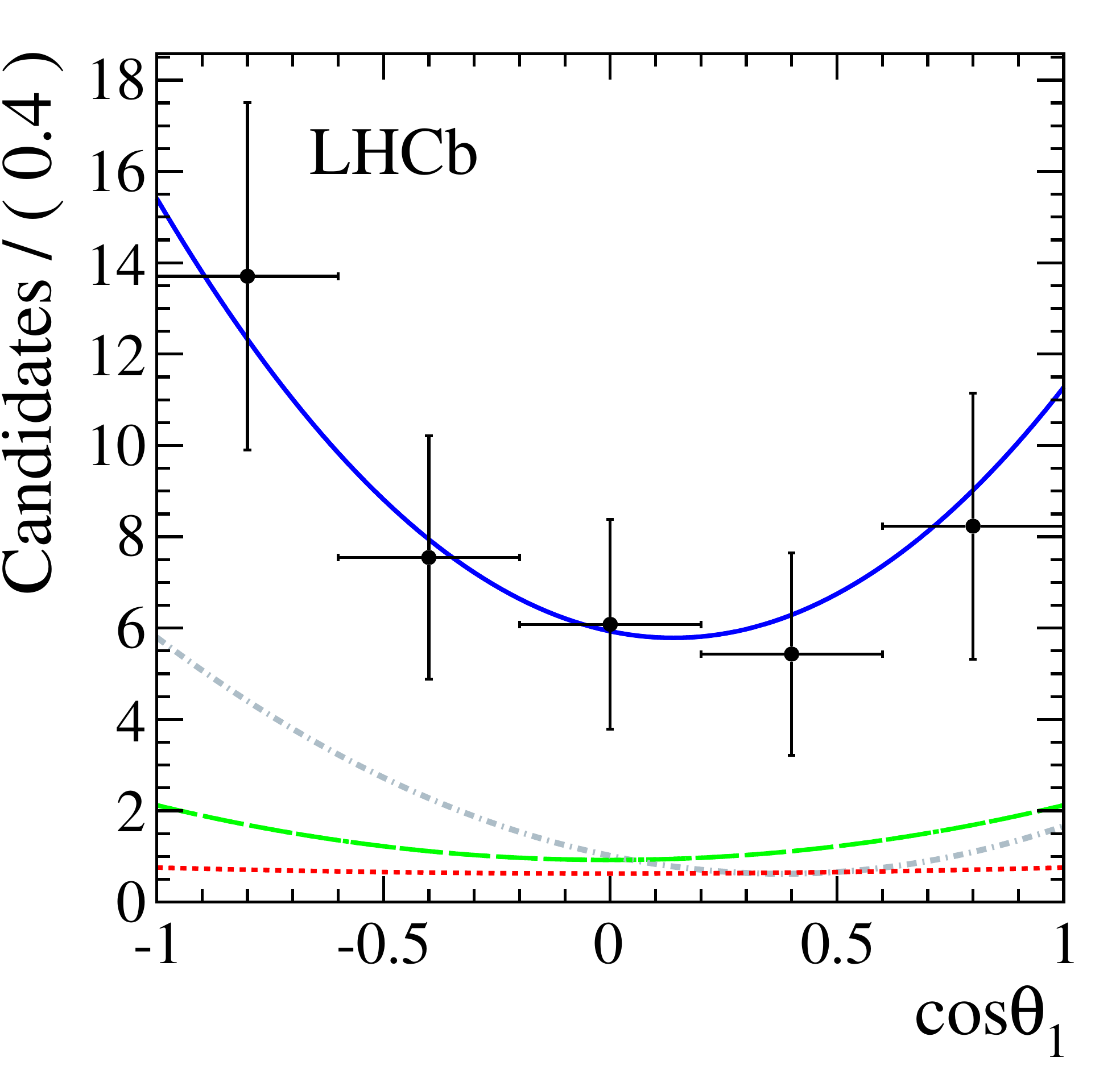}
    \includegraphics[scale=0.35]{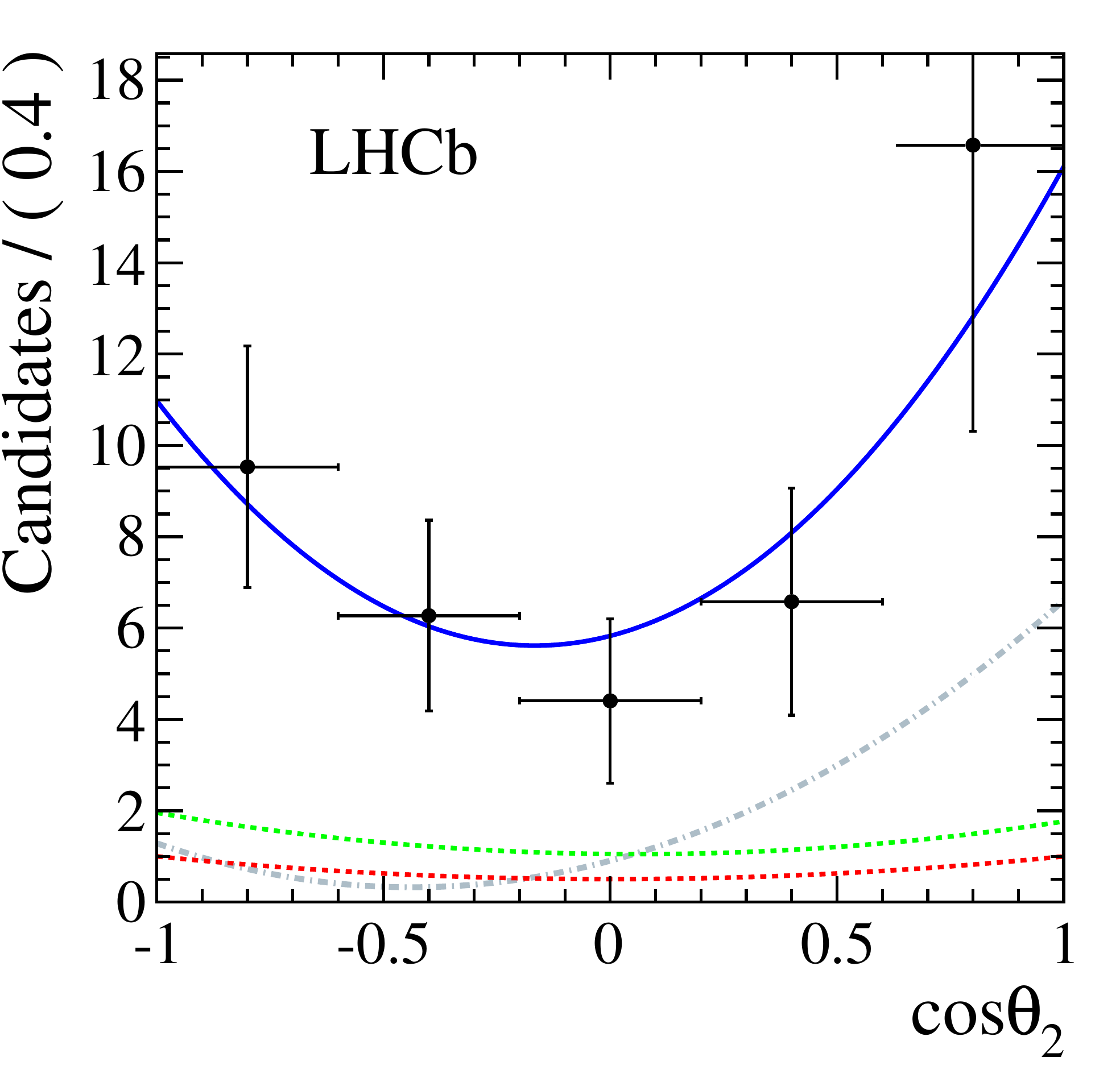}
  \caption{\small Result of the fit to the angular distribution of the \BsPhiKst
candidates in (left) $\cos\theta_1$
and (right) $\cos\theta_2$. The red dotted line corresponds
to the combinatorial background under the \Bs signal, the green dashed
line is the \BdPhiKst signal in the \Bs region and the grey
dotted-dashed line corresponds to the sum of the \swave and the interference terms.}
\label{fig:anglesBs}
\end{figure}

\noindent The different contributions to the systematic uncertainty are given in Table~\ref{tab:ang_syst}
and the one-dimensional projections of the angular distributions are shown Fig.~\ref{fig:anglesBs}.
Other possible systematic sources, such as the uncertainty in the
polarization parameters of the {\BdPhiKst}, are found to be negligible.

Considering all the above, the values obtained are
\begin{align}
     f_0 &= \phantom{-}0.51 \pm 0.15\,\stat \pm 0.07\,\syst, \nonumber \\
     f_\parallel &= \phantom{-}0.21 \pm 0.11\,\stat \pm 0.02\,\syst,  \nonumber \\
     \cos\delta_{\parallel} &= -0.18 \pm 0.52\,\stat \pm 0.29\,\syst. \nonumber
\end{align}

\noindent These results for the \BsPhiKst decay are consistent with the values measured in \BdPhiKst
decays of $f_0 = 0.494 \pm 0.036$, $f_\parallel = 0.212 \pm 0.035$
and $\cos\delta_{\parallel} = -0.74 \pm 0.10$~\cite{BabarPhiKst2008}.

\begin{table}[t]
\centering
\caption{Systematic uncertainties of the angular parameters.}
\begin{tabular}{cccc}
\hline
\hline
 Effect & $\Delta f_0$  & $\Delta f_\parallel$ & $\Delta \cos\delta_{\parallel}$ \\
\hline
 \swave & $0.07\phantom{0}$ & $0.02\phantom{0}$  &  $0.29\phantom{0}$ \\
Acceptance & $0.007$ & $0.005$ & $0.002$ \\
Combinatorial background & $0.02\phantom{0}$ & $0.01\phantom{0}$  &  $0.01\phantom{0}$ \\
\hline
Total & $0.07\phantom{0}$ & $0.02\phantom{0}$ & $0.29\phantom{0}$ \\
\hline
\hline
\end{tabular}
\label{tab:ang_syst}
\end{table}

\section{Summary and conclusions}
\label{sec:Conclusions}

A total of $30 \pm 6$ $\Bs \to (\KK)(\Kpi)$
candidates have been observed within the mass windows $1012.5 < M(\KK) < 1026.5\mevcc$ and 
$746 < M(\Kpi)< 1046\mevcc$. The result translates into a significance of
$6.2\,\sigma$. The analysis of the \KK and the \Kpi mass distributions
is consistent with $(84 \pm 2) \%$ of the signal originating from resonant
$\phi$ and \Kstarzb mesons. The significance of the \BsPhiKst resonant contribution is calculated
to be $6.1\,\sigma$.
The branching fraction of the decay is measured to be
\[
\BF(\BsPhiKst) = \left(1.10 \pm 0.24\,\stat \pm 0.14\,\syst \pm 0.08\left(\frac{f_d}{f_s}\right)\right) \times 10^{-6},
\]

\noindent using the \BdPhiKst decay as a normalization channel.
This result is roughly three times the theoretical expectation
in QCD factorization of $(0.4 \, {}^{+0.5}_{-0.3}) \times 10^{-6}$~\cite{Benm:2007rf}
and larger than the perturbative QCD value of ${(0.65 \, {}^{+0.33}_{-0.23}) \times 10^{-6}}$~\cite{Ali:2007ff},
although the values are compatible within $1\,\sigma$. The result is also higher
than the expectation of $\BF(\BdPhiKst) \times |V_{td}|^2/|V_{ts}|^2$.
Better precision on both the theoretical and experimental values would
allow this channel to serve as a probe for physics beyond the SM.

An angular analysis of the \BsPhiKst decay results in the
polarization fractions and phase difference
\begin{align}
     f_0 &= \phantom{-}0.51 \pm 0.15\,\stat \pm 0.07\,\syst, \nonumber \\
     f_\parallel &= \phantom{-}0.21 \pm 0.11\,\stat \pm 0.02\,\syst,  \nonumber \\
     \cos\delta_{\parallel} &= -0.18 \pm 0.52\,\stat \pm 0.29\,\syst. \nonumber
\end{align}

\noindent
The small value obtained for the longitudinal polarization fraction follows the trend
of the $b \to s$ penguin decays \BdPhiKst, \BsKstKst and \BsPhiPhi. The comparison
with the decay \BdKstKst, where $f_0 = 0.80^{+0.12}_{-0.13}$ ~\cite{PhysRevLett.100.081801},
shows a $2\,\sigma$ discrepancy. This is very interesting since the
loop-mediated amplitudes of each decay differ only in the flavour of the spectator quark.
The result is also compatible with the longitudinal polarization fraction $f_0= 0.40\pm 0.14$
measured in $\Bd \to \rho^0 \Kstarz$ decays~\cite{Lees:2011dq}, the penguin amplitude of which
is related to \BsPhiKst by $d \leftrightarrow s$ exchange. Finally, the result is smaller
than the prediction of perturbative QCD, $f_0 = 0.712 \, {}^{+0.042}_{-0.048}$, given in Ref.~\cite{Ali:2007ff}.

\section*{Acknowledgements}

\noindent We express our gratitude to our colleagues in the CERN
accelerator departments for the excellent performance of the LHC. We
thank the technical and administrative staff at the LHCb
institutes. We acknowledge support from CERN and from the national
agencies: CAPES, CNPq, FAPERJ and FINEP (Brazil); NSFC (China);
CNRS/IN2P3 and Region Auvergne (France); BMBF, DFG, HGF and MPG
(Germany); SFI (Ireland); INFN (Italy); FOM and NWO (The Netherlands);
SCSR (Poland); ANCS/IFA (Romania); MinES, Rosatom, RFBR and NRC
``Kurchatov Institute'' (Russia); MinECo, XuntaGal and GENCAT (Spain);
SNSF and SER (Switzerland); NAS Ukraine (Ukraine); STFC (United
Kingdom); NSF (USA). We also acknowledge the support received from the
ERC under FP7. The Tier1 computing centres are supported by IN2P3
(France), KIT and BMBF (Germany), INFN (Italy), NWO and SURF (The
Netherlands), PIC (Spain), GridPP (United Kingdom). We are thankful
for the computing resources put at our disposal by Yandex LLC
(Russia), as well as to the communities behind the multiple open
source software packages that we depend on.

\addcontentsline{toc}{section}{References}
\bibliographystyle{LHCb}
\bibliography{main}

\end{document}